\documentclass{article}
\usepackage[super,compress]{cite}
\usepackage{graphicx}

 \newcommand{\be}{\begin{equation}}
 \newcommand{\ee}{\end{equation}}
 \newcommand{\ben}{\begin{eqnarray}}
 \newcommand{\een}{\end{eqnarray}}
  \def\no{\nonumber}
\def\lb{\label}

  \usepackage[dvipdf]{epsfig}
\usepackage{color}
\begin{document}

%
%

\title{Exit from accelerated regimes by symmetry breaking in a universe
filled with fermionic and bosonic sources}

\author{Marlos O. Ribas\footnote{gravitam@yahoo.com},$\,$ Pedro  Zambianchi Jr.\footnote{zambianchi@utfpr.edu.br}
\\Departamento de F\'{\i}sica, Universidade Tecnol\'ogica Federal do Paran\'a\\
 Curitiba, Brazil
\\
Fernando P. Devecchi\footnote{devecchi@fisica.ufpr.br},$\,$ Gilberto M. Kremer\footnote{kremer@fisica.ufpr.br}\\
Departamento de F\'{\i}sica,
Universidade Federal do Paran\'a\\ Curitiba, Brazil
}

\maketitle


\begin{abstract}
In this work  we  investigate a universe filled with a fermionic field
  and  a  complex scalar field,  exchanging energy through a Yukawa
potential; the model encodes a symmetry breaking mechanism (on the bosonic sector). In a first case, when the mechanism is
not included,  the cosmological model furnishes a pure accelerated regime. In a second
case,  when including  the symmetry breaking mechanism, we verify that  the fermion
and one of the bosons, of Higgs type,
become massive, while the other boson is massless. Besides, the mechanism
shows to be  responsible for a
transition from an accelerated  to a decelerated
regime, which certifies the importance, in cosmological terms,  of its
role. After symmetry breaking, the total pressure of the fields
change its sign from negative to positive  corresponding to the
accelerated-decelerated transition. For
large times the universe becomes a dust (pressureless) dominated Universe.

\end{abstract}


\section{Introduction}

Cosmological models that include   fermionic sources are candidates
for explaining   regimes  of positive acceleration, and they  have been
studied recently in the works
\cite{1a,1b,1c,1d,2,3a,3b,3c,3d,3e}. Following  in that direction one possibility  is, in
particular,  to consider these fermionic fields as gravitational
sources for early  universes. These cases have been investigated by using
several  approaches, with results including exact solutions,
anisotropy-to-isotropy scenarios and self-interacting potentials (see for
example~\cite{1a,1b,1c,1d,4,3a,3b,3c,3d,3e}).

 An interesting question would be to investigate if it is
possible to implement a symmetry breaking
mechanism \cite{Ryder1,Ryder2,Ryder3,Ryder4},   including fermionic sources in a cosmological context. Furthermore, it would
be important to verify if this mechanism permit the entrance into a decelerated period dominated (for
instance)  by a bosonic constituent.
Taking these ideas into account, in this work we  propose
the description  of a  universe leaving an initial accelerated period,
including two constituents: a massless Dirac fermion  and a  complex scalar boson (also
massless), interacting, as in our previous effort\cite{RKD2},  through a
  Yukawa potential.  The  fundamental point here is that the model
lagrangian encodes, besides the diffeomorphism local
  invariance\cite{Wein}, typical of gravitational theories, a global symmetry controlled by the
group structure U(1) \cite{Ryder1,Ryder2,Ryder3,Ryder4};   of which the complex scalar
  field, mentioned above,  would be a representation.
Another fundamental fact is that the symmetry breaking mechanism
\cite{Ryder1,Ryder2,Ryder3,Ryder4}  associated to the U(1) invariance would be responsible for
  the generation of mass in this universe, and   the symmetry breaking
condition would be represented  by this  system reaching the so-called false-vacuum  state
\cite{8a,8b,8c}.  As expected, the false vacuum   is in fact  a continuum of
vacua. The permanent universe expansion implies  into the system
choosing  one of these vacua, and the U(1) symmetry is therefore
broken \cite{Ryder1,Ryder2,Ryder3,Ryder4}.   As far as this state
must be related to   {\it observables} the symmetry charge operator acting
on the vacuum state cannot give a different from
  zero result \cite{Ryder1,Ryder2,Ryder3,Ryder4}. So, speaking in terms of the lagrangian
formalism,   the field variables must be redefined
  in order to have a well-structured
lowest-energy state \cite{Ryder1,Ryder2,Ryder3,Ryder4}. This new lagrangian  density would be the
starting point for the dynamics after the
symmetry breaking, and, under proper conditions,  would describe the
universe entering into  a permanently
decelerated period.

Therefore, in few words,  we want to focus on
the symmetry breaking mechanism and  want to investigate what kind of role
this mechanism precisely  plays in an initially accelerating universe.
First, it is shown that without the symmetry
breaking mechanism the cosmological model furnishes a
pure accelerated regime. Second, when the symmetry is broken the fermion
and one of the bosons become massive, while the other boson remains
massless. The massive boson is identified as a Higgs-type  boson, and the
massless as a Goldstone boson \cite{Ryder1,Ryder2,Ryder3,Ryder4}. Then,  the symmetry breaking
mechanism  would be answering for the
accelerated-decelerated transition.

The work is structured as follows: in Section II we present the action
which includes a massless and non-self-interacting Dirac field and a
massless complex scalar field with a self-interacting potential. These
two fields interacts through a Yukawa-type potential. The analysis of
universe evolution without the symmetry breaking mechanism is given in
Section III, while in Section IV we introduce the Lagrangian and the
corresponding field equations after the symmetry breaking. In Section V we
show a numerical solution of the field equations with the
accelerated-decelerated transition and  the evolution of the total
pressure of the sources of the gravitational field. We state the main
conclusion of the work in Section VI. The metric signature used is
$(+,-,-,-)$ and units have been chosen so that $8\pi G=c=\hbar=1$.

\section{The action}

We are interested in investigating a cosmological model  described by a scalar and a  Dirac field. The scalar field $\phi$ is complex, massless with a self-interacting potential. The Dirac field $\psi$ is also massless and interacts with the complex scalar field through a Yukawa potential. The action is invariant with respect to a global symmetry controlled by the group structure U(1) and reads
\ben\no
S=\int\sqrt{-g}\,d^4x\bigg[\frac{1}{2}R+\frac{1}{2}\partial_\mu\phi^*\partial^\mu\phi-V(\phi^*\phi)+Y(\overline\psi\psi,\phi^*\phi)
\\\lb{1}
+\frac{i}{2}\left(\overline\psi\Gamma^\mu D_\mu\psi-D_\mu\overline\psi\Gamma^\mu\psi\right)\bigg].
\een
Here  $R$ denotes the scalar curvature, while $\psi$ and $\overline\psi=\psi^\dag\gamma^0$  are the spinor field and its adjoint, respectively. According to the general covariance principle,  the $4\times4$ Dirac-Pauli matrices
\ben
 \gamma^0=\left(
            \begin{array}{cc}
              1 & 0 \\
              0 & -1 \\
            \end{array}
          \right),\qquad \gamma^i=\left(
                           \begin{array}{cc}
                             0 & \sigma^i \\
                             -\sigma^i & 0 \\
                           \end{array}
                         \right),
 \een
  are replaced by their generalized versions $\Gamma^\mu=e^\mu_a\gamma^a$, where $\sigma^i$ are the $2\times2$ Pauli matrices  and $e^\mu_a$  the tetrad fields. Furthermore, the $\Gamma^\mu$ matrices satisfy  the generalized Clifford algebra $\{\Gamma^\mu,\Gamma^\nu\}=2g^{\mu\nu}$. The covariant derivatives in (\ref{1}) read
\ben\label{2}
D_\mu\psi= \partial_\mu\psi-\Omega_\mu\psi,\qquad
D_\mu\overline\psi=\partial_\mu\overline\psi+\overline\psi\Omega_\mu,
\een
where $\Omega_\mu$ is  the spin connection
\begin{equation}
\Omega_\mu=-\frac{1}{4}g_{\rho\sigma}[\Gamma^\rho_{\mu\delta}
-e_b^\rho(\partial_\mu e_\delta^b)]\Gamma^\delta\Gamma^\sigma,
\label{3}
\end{equation}
and $\Gamma^\nu_{\sigma\lambda}$  the Christoffel symbols.

The interaction between the fields is given by the Yukawa potential $Y(\overline\psi\psi,\phi^*\phi)=-\lambda\overline\psi\psi\phi^*\phi$ with $\lambda$ denoting a constant. Furthemore,  the self-interacting potential of the complex scalar field reads
\ben\lb{4}
V(\phi^*\phi)=-\frac{1}{2}\mu^2\phi^*\phi+\frac{1}{4}\xi^2(\phi^*\phi)^2,
\een
where $\mu$ and $\xi$ are constants.

\section{Before symmetry breaking}

The determination of the field equations are obtained from the variation of the action (\ref{1}) with respect the fields.

From the variation of (\ref{1}) with respect to the scalar fields $\phi$ and
$\phi^*$ we get the Klein-Gordon equations
\ben\lb{5a}
\nabla_\mu\nabla^\mu\phi-\mu^2\phi+\xi^2(\phi\phi^*)\phi+2\lambda\overline\psi\psi\phi=0,\\\lb{5b}
\nabla_\mu\nabla^\mu\phi^*-\mu^2\phi^*+\xi^2(\phi\phi^*)\phi^*+2\lambda\overline\psi\psi\phi^*=0.
\een

Dirac's equations are obtained from the variation of the action wit respect to $\overline\psi$ and $\psi$, yielding,
\ben\lb{6}
i\Gamma^\mu D_\mu\psi-\lambda\phi\phi^*\psi=0,\qquad i D_\mu\overline\psi\Gamma^\mu+\lambda\overline\psi\phi\phi^*=0.
\een
Furthermore, the left multiplication of the first equation by $\overline\psi$ summed with the right multiplication of the second one by $\psi$ leads to
\ben\lb{6a}
\overline\psi\Gamma^\mu D_\mu\psi+ D_\mu\overline\psi\Gamma^\mu\psi=0.
\een

The variation of (\ref{1}) with respect to the tetrad lead to Einstein's field equations
\ben\lb{7}
R_{\mu\nu}-\frac{1}{2}g_{\mu\nu}R=-T_{\mu\nu},
\een
where the energy-momentum tensor of the sources of the gravitational field reads
\ben\no
T^{\mu\nu}=\frac{i}{4}\left[\overline\psi\Gamma^\mu D^\nu\psi+\overline\psi\Gamma^\nu D^\mu\psi-D^\nu\overline\psi\Gamma^\mu\psi-D^\mu\overline\psi\Gamma^\nu\psi\right]+
\frac{1}{2}\left[\partial^\mu\phi^*\partial^\nu\phi\right.
\\\no
\left.+\partial^\nu\phi^*\partial^\mu\phi\right]
-g^{\mu\nu}\bigg[\frac{1}{2}\partial^\rho\phi^*\partial_\rho\phi-V(\phi^*\phi)+Y(\overline\psi\psi,\phi^*\phi)
\\\lb{8}
+\frac{i}{2}\left(\overline\psi\Gamma^\lambda D_\lambda\psi-D_\lambda\overline\psi\Gamma^\lambda\psi\right)\bigg].
\een

Let us consider  the case of a homogeneous and isotropic Universe spatially flat, described by the  Friedmann-Robertson-Walker (FRW) metric
\ben\lb{10}
ds^2=dt^2-a(t)^2(dx^2+dy^2+dz^2)
\een
where $a(t)$ denotes the cosmological scale factor. In this case the  Dirac-Pauli matrices and spin connection become
\ben\lb{11}
\Gamma^0=\gamma^0,\qquad
\Gamma^i=\frac{1}{a(t)}\gamma^i, \qquad \Omega_0=0, \qquad\Omega_i=\frac{1}{2}\dot a(t)\gamma^i\gamma^0,
\een
with the dot denoting time derivative.

For the FRW metric the acceleration equation that follows from Einstein's field equations (\ref{7}) reads
\ben\lb{12}
\frac{\ddot a}{a}=-\frac{1}{6}\left(\rho+3p\right),
\een
where  $\rho$ is the the energy density and $p$ the pressure $p$. Their expressions can be obtained from the energy-momentum tensor (\ref{8}) and are given by
\ben\lb{13a}
&&\rho=\frac{1}{2}\dot\phi^*\dot\phi+V(\phi^*\phi)-Y(\overline\psi\psi,\phi^*\phi),
 \\ \lb{13b}
&& p=\frac{1}{2}\dot\phi^*\dot\phi-V(\phi^*\phi).
 \een

 When the kinetic term  is much smaller than the potential one $\dot\phi^*\dot\phi/2\ll V(\phi^*\phi)$, the following relationship between the pressure and energy density can be established:
 \ben\lb{14}
 p=-\rho-Y(\overline\psi\psi,\phi^*\phi),
 \een

 Let us analyze the sign of the acceleration equation (\ref{12}) when the condition $\dot\phi^*\dot\phi/2\ll V(\phi^*\phi)$ is satisfied. In this case
 \ben\lb{15}
 \rho+3p=-2V(\phi^*\phi)-Y(\overline\psi\psi,\phi^*\phi).
 \een
 According to the weak energy condition (see e.g.~\cite{Wald}) $\rho+p\geq0$ and $\rho\geq0$, so that we have
 \ben
Y(\overline\psi\psi,\phi^*\phi)=-\lambda\overline\psi\psi\phi^*\phi\leq0,\qquad V(\phi^*\phi)\geq Y(\overline\psi\psi,\phi^*\phi),
 \een
 respectively.  From the first equation above we conclude that $\lambda>0$, provided that $\overline\psi\psi\phi^*\phi$ is a positive real number. The second one
  implies that $\rho+3p<0$ which together with (\ref{12}) implies that $\ddot a>0$, i.e.,  we are in the presence of an  accelerated regime .

 Hence under the condition $\dot\phi^*\dot\phi/2\ll V(\phi^*\phi)$ we have a pure accelerated regime and we need a mechanism to generate masses in order to bring the Universe to  a decelerated regime.
 In other words, the cosmological solutions of the field equations permit  an eternal expansion with
the energy density of the field constituents reaching a critical value at some point in the evolution. This critical values
are related to  the vacuum state of the scalar field and  this  promotes breaking of the U(1) symmetry.  As far as this state
must be related to an  observable  the field variables must be redefined in order to have a well-structured
lower-energy state.

Through the symmetry breaking process the fermion and the new boson  acquire masses, as expected (see e.g. \cite{Ryder1,Ryder2,Ryder3,Ryder4}).
This is to be considered  the starting point of a new period in the evolution of the Universe, as far as the
symmetry breaking  permits the exit from an accelerated period and the entrance on a decelerated era, which is a characteristic of a matter dominated universe.
It is important to reinforce here that the original Lagrangian does not  show a decelerated period among its solutions.
In fact, the end of accelerated period is possible only because of the symmetry breaking process, translated by the
use of new field dynamics controlled by a new Lagrangian which will be analyzed in the next section.

\section{After symmetry breaking}

A fundamental point here is that the Lagrangian encodes, besides the local diffeomorphism  invariance,
 a global symmetry controlled by the group structure U(1). This property will
be crucial  for our purposes,  since a symmetry breaking mechanism associated to the U(1) invariance would be responsible for the
generation of mass in this universe.

For the action (\ref{1}) we shall use the Higgs-Goldstone mechanism for mass generation of the fermions and bosons fields, from the choice of a stable potential minimum during its decay, breaking the global symmetry of the system. The fields acquire masses with the breaking symmetry mechanism, that is, we expand the fields around the potential minimum. The action (\ref{1}) is invariant with respect to the transformations of the global group U(1), characterized by $\phi\longrightarrow\phi e^{i\chi}$, where $\chi$ is global parameter that does not depend on the local space-time.

The minimum of the field does not occur for $\phi=0$ and  the symmetry breaking condition is represented  by the system reaching the
so-called false-vacuum  state  by imposing that ${dV}/{d\phi}=0$. This condition implies that  the components of the complex scalar field are given by
\begin{equation}\lb{16}
\phi_{1(0)}=\frac{\mu}{\xi}\cos\theta,\qquad \phi_{2(0)}=\frac{\mu}{\xi}\sin\theta.
\end{equation}
Furthermore,   as vacuum condition, we impose
\begin{equation}\lb{17}
\langle 0\mid \phi\mid 0\rangle =\frac{\mu}{\xi}e^{i\theta},
\end{equation}
where $\theta$ is the angle that defines a particular minimum.
As expected, the false vacuum   is in fact  a continuum of vacua,  labelled by the
 $\theta$ parameter of  expression (\ref{17}). The permanent universe expansion implies  into the system
choosing  one of these vacua, and the U(1) symmetry is therefore broken.
We stress  that
$\phi\mid0\rangle$ is not orthogonal to the vacuum state, making difficult to give an interpretation in terms of
particles due to the fact that in this case  the state $\phi\mid0\rangle$ would had  a non-zero number of particles,
and besides that it is not orthogonal to the vacuum state. As far as this state
must be related to an   observable  the field variables must be redefined in order to have a well-structured
lowest-energy state. The new fields are defined by the following equations:
\begin{equation}\lb{18}
\Phi=\phi-\langle 0\mid \phi\mid 0\rangle,
\end{equation}
with $\Phi=\varphi+i\eta$ where $\varphi$ and  $\eta$ are real scalar fields.
With these new variables we recover
the correct particle interpretation because of the condition $\langle 0\mid \Phi\mid 0\rangle=0,$.
Expanding the action (\ref{1}) in terms of the new fields
\ben\lb{19}
\phi=\varphi+i\eta+\frac{\mu}{\xi}e^{i\theta},\qquad \phi^*=\varphi-i\eta+\frac{\mu}{\xi}e^{-i\theta}
\een
we obtain a new action with U(1) symmetry broken. This new action describes a distinct cosmological model of the previous section, since apart from the massive fields we have new types of interaction vertices, as we shall show in the sequence.

The modifications which are introduced in the action (\ref{1}) are:
\ben\lb{20a}
\partial_\mu\phi^*\partial^\mu\phi=\frac{1}{2}\partial_\mu\varphi\partial^\mu\varphi+\frac{1}{2}\partial_\mu\eta\partial^\mu\eta,
\\\no
V(\phi^*\phi)=\mu\xi\eta^3\sin\theta+\mu\xi\eta\varphi^2\sin\theta+2\mu^2\eta\varphi \cos\theta\sin\theta+\mu\xi\varphi\eta^2\cos\theta
\\\lb{20b}
+\mu\xi\varphi^3\cos\theta+\mu^2\varphi^2\cos^2\theta-\frac{\mu^4}{4\xi^2}
+\frac{1}{2}\xi^2\eta^2\varphi^2+\mu^2\eta^2\sin^2\theta+\frac{1}{4}\xi^2\left(\eta^4+\varphi^4\right),
\\\lb{20c}
Y(\overline\psi\psi,\phi^*\phi)=-\lambda\overline\psi\psi\left(\varphi^2+\eta^2+2\varphi\frac{\mu}{\xi}\cos\theta
+2\eta\frac{\mu}{\xi}\sin\theta+\frac{\mu^2}{\xi^2}\right).
\een

Given the symmetry of the potential, the choice of the angle θ which leads to the new minimum of the potential should be immaterial. On the other hand, for a general $\theta$ the Lagrangian above apparently contains two massive bosonic fields, $\phi$ and $\eta$, thus, violating Goldstone theorem (see e.g. \cite{Ryder1,Ryder2,Ryder3,Ryder4}). It should be noted, however, that any interacting terms $\phi\eta$ renders the mass matrix off-diagonal and the masses of the free bosonic fields can no longer be identified. Only when the mass matrix is diagonal ($\theta = \pi/2$ or $\theta = 3\pi/2$) can the Lagrangian contain the physical fields. Since the vacuum state is degenerated after the symmetry breaking the system choose one of these states. Following the usual choice of the standard model we fix the vacuum state by considering $\theta=\pi/2$.

With the above choice the action (\ref{1}) becomes
\ben\no
S=\int\sqrt{-g}\,d^4x\bigg[\frac{1}{2}R+\frac{1}{2}\partial_\mu\eta\partial^\mu\eta-m_b\eta^2+\frac{i}{2}\left(\overline\psi\Gamma^\mu D_\mu\psi-D_\mu\overline\psi\Gamma^\mu\psi\right)
\\\lb{21}
+\frac{1}{2}\partial_\mu\varphi\partial^\mu\varphi
-m_f\overline\psi\psi-\widetilde V(\eta,\varphi)+\widetilde Y(\overline\psi\psi,\eta,\varphi)\bigg],
\een
where have introduced the modified potentials
\ben\lb{22a}
\widetilde V(\eta,\varphi)=\left(\eta^2+\varphi^2\right)\left[\mu\xi\eta+\frac{1}{4}\xi^2\left(\eta^2+\varphi^2\right)\right]-\frac{\mu^4}{4\xi^2},
\\\lb{22b}
\widetilde Y(\overline\psi\psi,\eta,\varphi)=-\lambda\overline\psi\psi\left(\varphi^2+\eta^2+2\frac{\mu}{\xi}\eta\right).
\een

From (\ref{21}) we infer that after symmetry breaking the Dirac field becomes massive due to its coupling with the scalar field through the Yukawa potential. Furthermore, the boson field $\eta$ also becomes  massive. The masses of the fermion and boson fields are given by $m_f=\lambda{\mu^2}/{\xi^2}$ and $m_b=\mu$, respectively. After the symmetry breaking process we have
a massless scalar field $\varphi$, whose quanta are known as Goldstone boson. Certainly these results agree with Goldstone theorem, which states that a continuous  symmetry breaking implies in the existence of a boson with vanishing mass, the so-called Goldstone boson. In the case of symmetry breaking in local groups in gauge theories, the same theorem is valid, however the field associated with the Goldstone boson can be absorbed by a gauge vectorial field by the choice of an adequate  gauge. In this manner the vectorial field acquires mass and gains one longitudinal degree of freedom. The massive boson cannot be eliminated by gauge choices and in the standard model of interactions it is know as Higgs boson and the mechanism associated with this process as Higgs mechanism.

It is worth to call attention to the fact that with the identification of the mass of the Dirac field, the modified Yukawa potential (\ref{22b}) is a real interaction potential, since it  is written in terms of products of the invariant of the spinors $\overline\psi\psi$ with the scalar fields $\eta$ and $\varphi$.

\section{Solution of the field equations}

In order to determine the cosmological solutions we have to solve the following system of coupled equations, which are written according to the FRW metric (\ref{10}):
\ben\lb{23a}
\ddot\eta+3\frac{\dot a}{a}\dot\eta+\mu\xi(3\eta^2+\varphi^2)+\xi^2\eta(\varphi^2+\eta^2)
+2\mu^2\eta-2\overline\psi\psi\left(\eta+\frac{\mu}{\xi}\right)=0,
\\\lb{23b}
\ddot\varphi+3\frac{\dot a}{a}\dot\varphi+2\mu\xi\eta\varphi+\xi^2\varphi(\eta^2+\varphi^2)-2\overline\psi\psi\varphi=0,
\\\lb{23c}
\dot{\overline\psi}\psi+\overline\psi\dot\psi+3\frac{\dot a}{a}\overline\psi\psi=0,
\\\lb{23d}
\frac{\ddot a}{a}=-\frac{1}{6}\left(\rho+3p\right),
\een
where the energy density $\rho$ and the pressure $p$ are given by
\ben\no
\rho=\frac{1}{2}\dot\eta^2+\frac{1}{2}\dot\varphi^2-\lambda \overline\psi\psi\left(\varphi^2+\eta^2+\frac{2\mu}{\xi}\eta+\frac{\mu^2}{\xi^2}\right)+\mu\xi\eta(\varphi^2+\eta^2)
\\\lb{24a}
+\mu^2\eta^2+\frac{1}{4}\xi^2\eta^2(\varphi^2+\eta^2)^2-\frac{\mu^4}{4\xi^2},
\\\lb{24b}
p=\frac{1}{2}\dot\eta^2+\frac{1}{2}\dot\varphi^2-\mu\xi\eta(\varphi^2+\eta^2)-\frac{1}{4}\xi^2\eta^2(\varphi^2+\eta^2)^2
-\mu^2\eta^2+\frac{\mu^4}{4\xi^2}.
\een
Equations (\ref{23a}), (\ref{23b}) and (\ref{23c}) follow from (\ref{5a}), (\ref{5b}) and (\ref{6a}), respectively.

From the coupled system of equations (\ref{23a}) -- (\ref{23d}) one may see that an algebraic solution of this system is very difficult.
The only equation that can be integrated is (\ref{23c}), whose solution is $\overline\psi\psi=C/a^3$, where $C$ is a constant of integration. Even the numerical solution is very complicated, since apart from the four constants $\lambda, \mu, \xi, C$ we need six initial conditions, namely, $(a(0),\dot a(0))$, $(\eta(0),\dot \eta(0))$ and $(\varphi(0),\dot\varphi(0))$. One of constants can be eliminated, because only the product $C\lambda$ appears in the equations. Moreover, we need only to specify the initial condition for $a(0)$, since the initial condition for $\dot a(0)$ can be obtained from the Friedmann equation
\ben
\dot a(0)=a(0)\sqrt{\frac{\rho(0)}{3}}.
\een

In the figures 1 and 2 the parameters chosen were  $C\lambda=1.5\times 10^{-5}$, $\mu=10^{-6}$, $\xi=0.17$ and the initial conditions $a(0)=1$, $\varphi(0)=0.25$, $\dot\varphi(0)=9\times 10^{-5}$, $\eta(0)=0.993$ and $\dot\eta(0)=3.4\times10^{-4}$.

\begin{figure}[h]
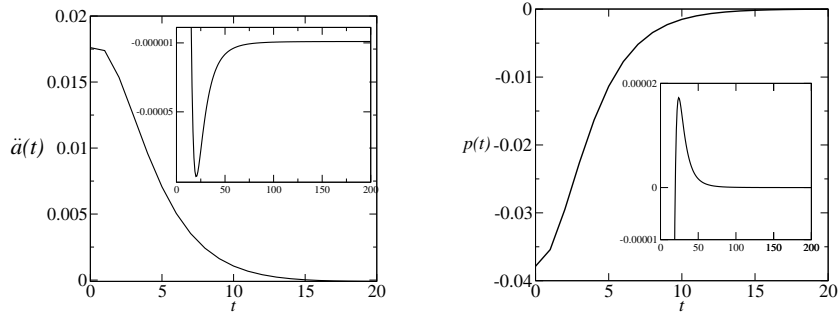
\vskip0.2cm
\centerline{\includegraphics[width=5cm]{fig1.eps}\hskip1cm \includegraphics[width=5cm]{fig2.eps}}
\caption{Left frame: Acceleration $\ddot a$ versus time $t$; right frame: pressure $p$ versus time $t$.}
\end{figure}

We observe from the left frame of Fig.1  that a strong accelerated regime for $0\leq t\approx 20$ is followed by a weak decelerated regime for $t>20$ that extends for a long  period of time (see the inset in this figure).  The transition from an accelerated  to a decelerated regime is a consequence of the symmetry breaking and the generation of two massive terms, a fermion and a boson. The right frame of Fig. 1 confirms also that this transition is connected to the fact that a large negative pressure that goes to zero for $0\leq t\approx 20$ is followed by a small positive pressure, for $t>20$, that tends to zero for long period of time (see the inset in this figure), which is a matter period dominated by dust.
From these figures we observe  a correspondence between the minimum of the deceleration with the maximum of the pressure.

\section{Conclusions}

In this work we have investigated  the symmetry breaking mechanism  associated to a global U(1) invariance, in a
universe filled with two sources: a Dirac fermionic field and a complex scalar field.  The global U(1) symmetry imply into a
family of degenerated vacua. The permanent universe
expansion encodes, among its dynamics, the system
choosing  one of these vacua, and the U(1) symmetry is therefore broken.

We have shown that without the symmetry breaking mechanism the cosmological model furnishes a pure accelerated regime.
However, due to the  symmetry breaking mechanism  one fermion and two bosons become the sources of this  universe.
The fermion and one of the bosons, of Higgs type, become massive, while the other boson is massless and
is interpreted as a Goldstone boson. This mechanism  is responsible not only for  generating masses, as expected, but  to allow a
transition from an accelerated regime to a decelerated
one. In this manner,  the coupling terms together with the massive
ones contribute to the end of an accelerated period,
after the breaking of the original global  U(1) symmetry.

Furthermore, we  verified that after symmetry breaking the total pressure of the fields
change its sign from negative to positive, which corresponds
to the  accelerated-decelerated transition. For
large times the deceleration of the universe is very slow
and the pressure tends to zero, i.e., it tends to a dust
dominating Universe.

In the future we are interested in investigating the consequences of a cosmological model with a local gauge symmetry, where the Goldstone mode could be absorbed by a vectorial field, through an adequate gauge fixing. In this case the massive scalar field will play the role of a Higgs-type boson.

\section*{Acknowledgments}

One of the authors (GMK) acknowledges the financial support of Conselho Nacional de Desenvolvimento
Cient\'{\i}fico e Tecnol\'ogico -- CNPq  (Brazil).

\end{document}